\documentclass[preprint,prc,showpacs,preprintnumbers,superscriptaddress,floatfix]{revtex4}
\usepackage{dcolumn}
\usepackage{bm}
\usepackage{longtable}
\usepackage{graphicx,epsfig,latexsym,amssymb}
\usepackage{multirow,amsmath,array,booktabs,color}
\usepackage[section]{placeins}

\begin{document}

\title{Resonant states of deformed nuclei in complex scaling method}
\author{Quan Liu}
\affiliation{School of Physics and Material Science, Anhui University, Hefei 230039,
P.R.China}
\author{Jian-You Guo}
\email{jianyou@ahu.edu.cn}
\affiliation{School of Physics and Material Science, Anhui University, Hefei 230039,
P.R.China}
\author{Zhong-Ming Niu}
\affiliation{School of Physics and Material Science, Anhui University, Hefei 230039,
P.R.China}
\author{Shou-Wan Chen}
\affiliation{School of Physics and Material Science, Anhui University, Hefei 230039,
P.R.China}
\date{\today }

\begin{abstract}
We develop a complex scaling method for describing the resonances of
deformed nuclei and present a theoretical formalism for the bound and
resonant states on the same footing. With $^{31}$Ne as an illustrated
example, we have demonstrated the utility and applicability of the extended
method and have calculated the energies and widths of low-lying neutron
resonances in $^{31}$Ne. The bound and resonant levels in the deformed potential
are in full agreement with those from the multichannel scattering approach.
The width of the two lowest-lying resonant states shows a novel evolution
with deformation and supports an explanation of the deformed halo for $^{31}$Ne.
\end{abstract}

\maketitle

\section{Introduction}

The investigation of continuum and resonant states is an important subject
in quantum physics. In recent years, there has been an increasing interest
in the exploration of nuclear single particle states in the continuum. The
construction of the radioactive ion beam facilities makes it possible to
study exotic nuclei with unusual N/Z ratios. In these nuclei, the Fermi
surface is usually close to the particle continuum, thus the contribution of
the continuum and/or resonances is essential for exotic nuclear phenomena~%
\cite{Dobaczewski96,Poeschl97,Meng06,Hamamoto10,Zhou10}.

Several techniques have been developed to study the resonant states in the
continuum. One of them is the R-matrix theory in which resonance parameters
(i.e. energy and width) can be reasonably determined from fitting the
available experimental data \cite{Wigner47}. The extended R-matrix theory
\cite{Hale87} and the K-matrix theory \cite{Humblet91} have been also
developed. The conventional scattering theory is also an efficient tool for
studying resonances. More precisely, the scattering phase shift method is
adopted and the resonant state is determined from the pole of the S matrix
\cite{Taylor72}. Computationally, it is desired to deduce the properties of
unbound states from the eigenvalues and eigenfunctions of the Hamiltonian
for bound states so that the methods developed for bound states can still be
used. For this purpose, bound-state-type methods have been developed,
including the real stabilization method (RSM) \cite{Hazi70}, the complex
scaling method (CSM) \cite{Ho83}, and the analytic continuation in the
coupling constant (ACCC) method \cite{Kukulin89}. These bound-state-like
methods have gained a great development. For examples, many efforts have
been made in order to calculate more efficiently resonance parameters with
the RSM~\cite{Taylor76,Mandelshtam94,Kruppa99}. Its extension to the
relativistic framework has been presentes in Ref. \cite{Zhang08}, where the
resonance parameters obtained are comparable with those from the
relativistic ACCC calculations. Combined with the cluster model, the ACCC
approach has been used to calculate the energies and widths of resonant
states in some light nuclei~\cite{Tanaka97,Tanaka99}. An attempt to extend
the ACCC approach to the relativistic mean field (RMF) model was made in
Ref.~\cite{Yang01}. Further development of the formalism was presented in
Ref.~\cite{Zhang04}, where the wave functions of the resonant states were
determined by the ACCC method. Based on the relativistic extension of the
ACCC method, the structure of resonant levels was investigated and good
pseudospin symmetry was disclosed in realistic nuclei~\cite{Guo05,Guo06}.
Although it involves the solution of a complex eigenvalue problem which
causes some difficulties in practice, the CSM has been widely and
successfully used to study resonances in atomic and molecular systems~\cite%
{Ho83,Moiseyev98} and atomic nuclei~\cite{Kruppa97,Arai06,Aoyama06,Myo12}.
Its generalization to the relativistic problem was first outlined by Weder
\cite{Weder74}. Recent progress includes the following: Alhaidari \cite%
{Alhaidari07} has presented a general and systematic development of an
algebraic extension of the CSM to the relativistic problem by expanding the
Dirac spinors in Laguerre functions , Bylicki et al. \cite{Bylicki08} have
formulated a positive-energy-space-projected relativistic Hylleraas
configuration-interaction method based on the complex coordinate rotation~,
and we have developed the CSM within the RMF framework in satisfactory
agreements with the RSM, the scattering phase-shift method and the ACCC
approach \cite{Guo10}.

All these studies are mainly for spherical nuclei. Recently, the resonances
of deformed nuclei have attracted additional attention. The interplay
between the deformation and the low-lying resonance is very interesting for
the open shell nucleus close to the drip line. The recent experimental data
on the Coulomb breakup of the nucleus $^{31}$Ne \cite{Nakamura09} have been
well interpreted in terms of deformation with particle populating in the
resonant levels \cite{Hamamoto10}. The deformed halos have been investigated
systemically in Ref. \cite{Zhou10}, where the resonances in the continuum
play an important role. Some techniques have been developed to investigate
the resonances of a deformed system. In Refs. \cite{Ferreira97,Yoshida05},
the resonances as a function of deformation in an axially deformed
Woods-Saxon potential and a quadrupole-deformed finite square-well potential
without a spin-orbit component have been studied by solving the
coupled-channel Schr\"{o}dinger equation. In Refs. \cite{Cattapan00,Hagino04}%
, the ACCC method and RSM have been extended to investigate the resonances
in axially deformed Woods-Saxon potentials. The single-particle resonant
states in an axially deformed Gaussian potential without spin and tensor
components have been investigated by the contour-deformation method in
momentum space \cite{Hagen06}. Using the multichannel scattering approach,
Hamamoto \cite{Hamamoto05} has studied how the single-particle energies
change from bound to resonant levels when the depth of the potential is
varied. This method has also been applied for the resonances of a Dirac
particle in a deformed potential \cite{Li10}. Hamamoto \cite{Hamamoto12} has
performed a systematic research on the resonances of deformed nuclei by the
multichannel scattering approach, which includes the evolution of shell
structure for exotic nuclei. However, as we know, CSM has not been used to
investigate the resonances of deformed nuclei. Due to the great success of
CSM in describing the resonances for a spherical system, in this paper, we
extend CSM to a deformed system, and examine its applicability and
efficiency for the resonances of deformed nuclei.

\section{Formalism}

Our purpose is to extend the CSM to describe the resonances of deformed
nuclei. This scheme has a certain advantage, i.e., the bound and resonant
states can be treated on the same footing, because the complex scaled
functions for the resonant states are square integrable just like those for
the bound states \cite{Moiseyev98}. In the following, we start our scenario
with the single-particle Hamiltonian give as
\begin{equation}
H=T+V\ ,  \label{Hamiltonian}
\end{equation}%
where $V$ is introduced with the axially symmetric quadruple-deformed
potential consisting of the following three parts:
\begin{eqnarray}
V_{\text{cent}}\left( r\right)  &=&V_{0}f\left( r\right) ,  \notag \\
V_{\text{cou}}\left( \vec{r}\right)  &=&-\beta _{2}V_{0}k\left( r\right)
Y_{20}\left( \vartheta ,\varphi \right) ,  \notag \\
V_{\text{so}}\left( r\right)  &=&-\frac{1}{2}vV_{0}g(r)\left( \vec{s}\cdot
\vec{l}\right) .  \label{potential}
\end{eqnarray}%
Here $k\left( r\right) =r\frac{df\left( r\right) }{dr}$ and $g(r)=\frac{%
\Lambda ^{2}}{r}\frac{df\left( r\right) }{dr}$. The parameter $\Lambda $ is
the reduced Compton wavelength of nucleon $\hslash /M_{r}c$. Similar to Ref.%
\cite{Hamamoto05}, a Woods-Saxon type potential is employed for $f\left(
r\right) =\frac{1}{1+\exp \left( \frac{r-R}{a}\right) }$. To explore the
resonant states, the CSM is used. A relative coordinate $r$ in Hamiltonian $H
$ and wave function $\psi $ is complex scaled as
\begin{equation}
U(\theta ):r\rightarrow re^{i\theta }.  \label{tranformation}
\end{equation}%
Then, the transformed Hamiltonian and wave function are defined as $%
H_{\theta }=U\left( \theta \right) HU\left( \theta \right) ^{-1}$ and $\psi
_{\theta }=U\left( \theta \right) \psi $, where $\psi _{\theta }$ is a
square integrable function. The corresponding complex scaled equation is
\begin{equation}
H_{\theta }\psi _{\theta }=E_{\theta }\psi _{\theta }.  \label{Complexeq}
\end{equation}%
From ABC theorem\cite{Aguilar71}, the following is known: (i) a bound-state
eigenvalue of $H$ is also an eigenvalue of $H_{\theta };$ (ii) a resonance
pole $E=E_{r}-i\Gamma /2$ of the Green operator of $H$ is an eigenvalue, $%
E_{\theta }=E_{r}-i\Gamma /2$ of $H_{\theta }$; and (iii) the continuous
part of the spectrum is rotated around the origin of the $E-$ plane by an
angle $2\theta $.

To solve the complex scaled equation (\ref{Complexeq}), it is convenient to
adopt the basis expansion method. For the axially symmetrically deformed
system, the parity $\pi =(-1)^{N}=(-1)^{l}$ and the projection of the total
angular momentum along the symmetry axis $\Omega $ are good quantum numbers;
$\psi _{\theta }$ can be expanded as
\begin{equation}
\psi _{\theta }=\sum_{i}c_{i}\left( \theta \right) \phi _{i}\text{ },
\label{Expand}
\end{equation}%
where $\phi _{i}=R_{nl}(r)Y_{lm_{l}}(\vartheta ,\varphi )\chi
_{m_{s}}(\sigma _{z})$ and the sum $i$ runs over the quantum numbers $n,l,$%
and $m_{l}$ with $\Omega =m_{l}+m_{s}$. $R_{nl}(r)$ is the radial function
of a spherical harmonic oscillator (HO) potential,
\begin{equation}
R_{nl}(r)=\frac{1}{b_{0}^{3/2}}\sqrt{\frac{2\left( n-1\right) !}{\Gamma
\left( n+l+1/2\right) }}x^{l}L_{n-1}^{l+1/2}\left( x^{2}\right) e^{-x^{2}/2}%
\text{, }n=1,2,3,\text{ }....  \label{hobasis}
\end{equation}%
$x=r/b_{0}$ is the radius measured in units of the oscillator length $b_{0}$%
. $Y_{lm_{l}}(\vartheta ,\varphi )$ is the spherical harmonics. $\chi
_{m_{s}}(\sigma _{z})$ represents the spin wave function. The upper limit of
the radial quantum number $n$ is determined by the corresponding major shell
quantum number $N=2\left( n-1\right) +l$. Inserting the ansatz (\ref{Expand}%
) into the complex scaled equation (\ref{Complexeq}) and using the
orthogonality of wave functions $\phi _{i}$ one arrives at a symmetric
matrix diagonalization problem,
\begin{equation}
\sum_{i}\left[ T_{i^{\prime },i}+V_{i^{\prime },i}\right] c_{i}=E_{\theta
}c_{i^{\prime }}\text{ },  \label{matrixeq}
\end{equation}%
where $T_{i^{\prime },i}$ and $V_{i^{\prime },i}$ are presented as%
\begin{eqnarray}
T_{i^{\prime },i} &=&e^{-i2\theta }\int \phi _{i^{\prime }}\left( -\frac{%
\hbar ^{2}}{2M}\left( \frac{d^{2}}{dr^{2}}+\frac{2}{r}\frac{d}{dr}\right) +%
\frac{\vec{l}^{2}}{2Mr^{2}}\right) \phi _{i}d\vec{r}\text{ },
\label{Tmatrix} \\
V_{i^{\prime },i} &=&\int \phi _{i^{\prime }}V\left( \vec{r}e^{i\theta
}\right) \phi _{i}d\vec{r}\text{ }.  \label{Vmatrix}
\end{eqnarray}%
Putting $\phi _{i}$ into the above expressions, the matrix elements $%
T_{i^{\prime },i}$ are obtained as
\begin{equation}
T_{i^{\prime },i}=e^{-i2\theta }\frac{\hbar ^{2}}{2Mb_{0}^{2}}\left[ \sqrt{%
n\left( n+l+1/2\right) }\delta _{n^{\prime },n+1}+(2n+l-1/2)\delta
_{n^{\prime },n}+\sqrt{\left( n-1\right) \left( n+l-1/2\right) }\delta
_{n^{\prime },n-1}\right] \delta _{l^{\prime }l}\delta _{m_{l}^{\prime
}m_{l}}\delta _{m_{s}^{\prime }m_{s}}  \label{Telement}
\end{equation}%
Similarly, the matrix elements $V_{i^{\prime },i}$ are obtained as
\begin{eqnarray}
V_{i^{\prime },i}^{\text{cent}} &=&\left\langle \phi _{i^{\prime
}}\right\vert V_{\text{cent}}(re^{i\theta })\left\vert \phi
_{i}\right\rangle =V_{0}\left\langle n^{\prime }l^{\prime }\right\vert
f(re^{i\theta })\left\vert nl\right\rangle \delta _{l^{\prime }l}\delta
_{m_{l}^{\prime }m_{l}}\delta _{m_{s}^{\prime }m_{s}}\text{ },  \label{Vcent}
\\
V_{i^{\prime },i}^{\text{cou}} &=&\left\langle \phi _{i^{\prime
}}\right\vert V_{\text{cou}}\left( \vec{r}e^{i\theta }\right) \left\vert
\phi _{i}\right\rangle =-\beta _{2}V_{0}\left\langle n^{\prime }l^{\prime
}\right\vert k\left( re^{i\theta }\right) \left\vert nl\right\rangle
\left\langle l^{\prime }m_{l}^{\prime }\right\vert Y_{20}\left( \vartheta
,\varphi \right) \left\vert lm_{l}\right\rangle \delta _{m_{s}^{\prime
}m_{s}}\text{ },  \label{Vcou} \\
V_{i^{\prime },i}^{\text{so}} &=&\left\langle \phi _{i^{\prime }}\right\vert
V_{\text{so}}\left\vert \phi _{i}\right\rangle =-\frac{1}{2}%
vV_{0}\left\langle n^{\prime }l^{\prime }\right\vert g(re^{i\theta
})\left\vert nl\right\rangle \left\langle l^{\prime }m_{l}^{\prime
}m_{s}^{\prime }\right\vert \left( \vec{s}\cdot \vec{l}\right) \left\vert
lm_{l}m_{s}\right\rangle \text{ }.  \label{Vso}
\end{eqnarray}%
In these expressions, $\left\langle l^{\prime }m_{l}^{\prime }\right\vert
Y_{20}\left( \vartheta ,\varphi \right) \left\vert lm_{l}\right\rangle $ and
$\left\langle l^{\prime }m_{l}^{\prime }m_{s}^{\prime }\right\vert \left(
\vec{s}\cdot \vec{l}\right) \left\vert lm_{l}m_{s}\right\rangle $ can be
calculated by the usual angular momentum formulas, and the radial parts of
integration can be calculated with the Gauss quadrature approximation. With
the matrix elements $T_{i^{\prime },i}$, $V_{i^{\prime },i}^{\text{cent}}$, $%
V_{i^{\prime },i}^{\text{cou}},$ and $V_{i^{\prime },i}^{\text{so}}$, the
solutions of the complex scaling equation (\ref{Complexeq}) can be obtained
by diagonalizing the matrix $H_{\theta }$. The eigenvalues of $H_{\theta }$
representing bound states or resonant states do not change with $\theta $,
while the eigenvalues representing the continuous spectrum rotate with $%
\theta $. The former are associated with resonance complex energies $%
E-i\Gamma /2$, where $E$ is the resonance position and $\Gamma $ is its
width.

\section{Numerical details and Results}

With the theoretical formalism, we explore the resonant states for the
realistic physical system. In the numerical details, the complex-scaled Schr%
\"{o}dinger equation is solved by expansion in the HO basis with $60$
oscillator shells. The oscillator frequency of the HO basis is fixed as $%
\hslash \omega _{0}=41A^{-1/3}$ MeV. For the comparison with the results
from the multichannel scattering approach \cite{Hamamoto05}, $^{31}$Ne is
taken as an example. The corresponding parameters are chosen to be same as
those in Ref.\cite{Hamamoto10}, i.e., $a=0.67$ fm, $V_{0}=-39$ MeV, $v=32$,
and the nuclear radius $R$ varying with the mass number $A$ of the system
with $R=r_{0}A^{1/3},$ where $r_{0}=1.27$ fm is used. To achieve the
information on the resonant states, we diagonalize $H_{\theta }$. The
eigenvalues of $H_{\theta }$ for these states with the quantum numbers
varying from $\Omega ^{\pi }=\frac{1}{2}^{\pm }$ to $\Omega ^{\pi }=\frac{9}{%
2}^{\pm }$ are plotted in Fig.1, where the rotation angle $\theta =18^{\circ
}$ and the quadruple deformation $\beta _{2}=0.1$ are adopted. All the
eigenvalues of $H_{\theta }$, which correspond to the bound, the resonant,
and the continuum, are respectively labeled as open squares, red open
circles, and open circles. The solid line marks the position of the
continuum with the rotation angle $2\theta $. From Fig.1, one sees clearly
that the eigenvalues of $H_{\theta }$ fall into three regions: the bound
states populate on the negative energy axis, while the continuous spectrum
of $H_{\theta }$ rotates clockwise with the angle $2\theta $, and resonances
in the lower half of the complex energy plane located in the sector bound by
the new rotated cut line and the positive energy axis get exposed and become
isolated. In realistic calculations, because the finite basis is used, the
continuous spectrum of $H_{\theta }$ consists of a string of points.

Next, we demonstrate how the resonant states become exposed by complex
rotation. The eigenvalues of $H_{\theta }$ with $\theta =8^{\circ
},12^{\circ },16^{\circ },$and $20^{\circ }$ are respectively displayed in
Figs.2(a)-2(d), where the other parameters are the same as those in Fig.1.
From Fig.2 one can see that, when $\theta =8^{\circ }$, the resonances still
have not been completely separated from the continuum, and to identify the
resonances in the continuum is still relatively difficult although several
points are marked as the resonances in the complex energy plane. When the
rotation angle is added to $\theta =12^{\circ }$, the resonant states are
clearly isolated from the continuum and distinguishing the resonances from
the continuum becomes easy. When the rotation angle is added to $\theta
=16^{\circ }$, the resonant states are completely separated from the
continuum, and clearly exposed in the complex energy plane. This is because
the continuous spectrum of $H_{\theta }$ rotates clockwise with ${\theta }$,
while the resonances are almost independent of the rotation angle $\theta $.
From Figs.2(c) and 2(d), the rotation angle is incresed from $\theta
=16^{\circ }$ to $\theta =20^{\circ }$, and the change of position of the
resonant states in the complex energy plane is almost not observable. This
indicates that the resonant states can be unambiguously determined at the
same time as long as the rotation angle is chosen to be large enough, which
is one of the advantages of CSM. In the present calculations, the exposure
of the resonant states relies on the scale of the rotation angle, while
their position in the complex energy plane remains almost unchanged with the
variation of $\theta $; i.e., the energies and widths are nearly $\theta $
independent, which also reflects that the stability of the results with
respect to the variation of $\theta $ is considerably satisfactory.

As we look forward to the resonances of the deformed nuclei states, it is
interesting to observe the movement of the resonant states in the position
with deformation in the complex energy plane, which is helpful to recognize
intuitively the resonances of the deformed nuclei. In Fig.3, we show the
eigenvalues of $H_{\theta }$ with deformation $\beta _{2}=0.0,-0.1,-0.2,$and
$-0.3$. The other parameters are the same as those in Fig.1. When the system
is spherically symmetrical ($\beta _{2}=0.0$), only three resonances are
observed, which correspond respectively to the resonant states $%
1f_{7/2},1f_{5/2}$, and $1g_{9/2}$. When the spherical symmetry of the
system is broken ($\beta _{2}\neq 0.0$), the 3 resonances are broken down
into 12 resonances, which correspond respectively to the resonant states
1/2[330], 3/2[321], 5/2[312], 7/2[303], 1/2[310], 3/2[301], 5/2[303],
1/2[440], 3/2[431], 5/2[422], 7/2[413], and 9/2[404] (denoted with the
asymptotic quantum numbers $\Omega \lbrack Nn_{z}\Lambda ])$. The location
of the 12 resonant states in the complex energy plane appears to be a
significant movement when the deformation is changed. With the development
of deformation towards the direction of larger deformation, the 12 resonant
states in the position are more clearly separated. The similar movement of
the position towards the prolate shapes has also been obtained, and not been
displayed here.

Although the resonances of several points of deformation are displayed in
Fig.3, the resonance parameters of these points have not been fully
determined for the resonant states exposed in Fig.1. In the following, we
check how the results depend on $\theta $ in the present calculations in
order to present the accurate values of the resonance parameters. According
to the mathematical theorem the resonance energies should be independent of $%
\theta $ (for large enough $\theta $). However, in the present numerical
approximation, resonance parameters are dependent slightly on $\theta $ due
to basis expansion with finite shells, there will be no eigenvector that is
completely independent of $\theta $. The resonance energies will move along
trajectories in the complex energy plane as a function of $\theta $. The
best estimate for a resonance energy is given by the parameter $\theta $
value for which the rate of change with respect to $\theta $ is minimal. To
locate the minimal rate point, we perform repeated diagonalizations of the
eigenvalue problem of $H_{\theta }$ with different $\theta $ values in every
deformation concerned. The illustrated results are shown in Fig.4 for the
resonant state 7/2[303] of the deformation point $\beta _{2}=0.3$, where $N$
represents the number of oscillator shells of the basis. From Fig.4 one sees
clearly that, for different $N$, the dependence of the resonance parameters
on $\theta $ shows a different trajectory, and the range of the trajectory
becomes smaller with the increasing $N$ when $N\geqslant 50$. In every $%
\theta $-trajectory, there exists a point corresponding to the minimal rate
of change of the resonance parameters with respect to $\theta $, which
presents the optimal value of the resonance parameters. Despite the
difference in the $\theta $ trajectory, the optimal value of the resonance
parameters tends to the same result with the increasing $N$, especially when
$N\geqslant 50$. Although the movement of the optimal value in the complex
energy surface is observable, that is because we display these results in a
small energy scale. Compared with the continuous spectra, the movement is
negligible, especially when $N\geqslant 50$. Hence, the size of the basis
being chosen as $N=60$ is good enough in the previous calculations. To more
accurately determine the value of the resonance parameters, a close up of
the $\theta $ trajectory is plotted in Fig.5 for $N=60$. In the vicinity of $%
\theta =17^{\circ }$, the resonance parameters are almost independent of $%
\theta $, i.e., $\frac{dE_{\theta }}{d\theta }\approx 0$, which presents the
optimal value of the resonance parameters.

By using the $\theta $ trajectory, the resonance parameters of any
deformation point can be determined for the resonant states exposed in
Fig.1. In Fig.6, we show the evolution of the single-particle energy with
the deformation $\beta _{2}$ for the bound and resonant states. In
comparison with Ref.\cite{Hamamoto10}, consistent results are obtained for
the bound levels and the low-lying resonant levels. In addition, several
higher-lying resonant levels are also obtained in the present calculations.
Especially, the width is obtained for all the resonant states. In Fig.7, we
plot the width for the four lower-lying resonant states 1/2[330], 3/2[321],
5/2[312], and 7/2[303]. For the resonant state 7/2[303], the resonance
energy and the width increase monotonously with the deformation varying from
$\beta _{2}=-0.4$ to $0.6$. For the resonant state 5/2[312], the energy and
the width display an almost horizontal change with the deformation. For the
resonant states 1/2[330] and 3/2[321], with the deformation varying from $%
\beta _{2}=-0.4$ to $0.6$, the energy decreases monotonously, while the
width shows a unusual trend. Close to $\beta _{2}=0.0$, the value of the
width appears to be at a minimum, which implies the spherical shape is more
stable for the two resonant states. After the minimum appears, the resonance
width increases with the increasing of deformation, which means these nuclei
states become even more unstable as their levels become lower. With a
continued increase in deformation, the width appears to be at a maximum
value. Therefore, it is relatively difficult for the halo to forming in the
deformation interval from $\beta _{2}=0.0$ to the position of the width
appearing at a maximum value for $^{31}$Ne. Thereafter, the resonance width
decreases with the increase in deformation. When the deformation increases
beyond $\beta _{2}=0.25$, the level 1/2[330] becomes weakly bound. In the
vicinity of $\beta _{2}=0.3$, the levels 1/2[330] and 3/2[202] appear to be
crossing. With a further increase in deformation, the energy gap between
1/2[330] and 3/2[202] gets larger and a maximum appears at $\beta _{2}=0.43$%
. Thus, in this deformation interval from $\beta _{2}=0.25$ to $\beta
_{2}=0.43$, the 21th neutron can occupy the weakly bound level 1/2[330] or
3/2[202], and the halo can be formed there. The most likely position for
forming a halo should be near $\beta _{2}=0.3$. This information indicates
that $^{31}$Ne may be a halo nucleus with deformation $\beta _{2}\sim 0.3$.

\section{Summary}

In summary, the complex scaling method is extended to describe the
resonances of deformed nuclei. A theoretical formalism is presented in which
the bound and resonant states are treated on the same footing. $^{31}$Ne is
chosen as an example; the utility and applicability of the extended method
are demonstrated. It is found that the exposure of resonant states relies on
the scale of the rotation angle, while their position in the complex energy
plane remains almost unchanged with the variation of $\theta $, which
reflects that the stability of the results with respect to the variation of $%
\theta $ is considerably satisfactory. The movement of the resonant states
in the position with deformation is exhibited in complex energy plane, where
it is seen that 3 degenerate resonances are separated into 12 resonances
with the development of deformation towards the oblate or prolate shapes.
The bound and resonant levels obtained for $^{31}$Ne are in agreement with
those from the coupling-channel calculations, where the resonances are
determined by a multichannel scattering approach. With the deformation
evolution from the oblate to prolate shapes, the bound and resonant levels
show a clear shell structure. Especially, the width of the two lowest-lying
resonant states shows a novel evolution with deformation, and supports an
explanation of the deformed halo for $^{31}$Ne.

\begin{acknowledgments}
Helpful discussions with Professor Hamamoto are acknowledged. This work was
partly supported by the National Natural Science Foundation of China under
Grants No. 10675001, No. 11175001, and No. 11205004; the Program for New
Century Excellent Talents in University of China under Grant No.
NCET-05-0558; the Excellent Talents Cultivation Foundation of Anhui Province
under Grant No.2007Z018; the Education Committee Foundation of Anhui
Province under Grant No. KJ2009A129; and the 211 Project of Anhui University.
\end{acknowledgments}

\begin{figure}[tbp]
\includegraphics[width=8.cm]{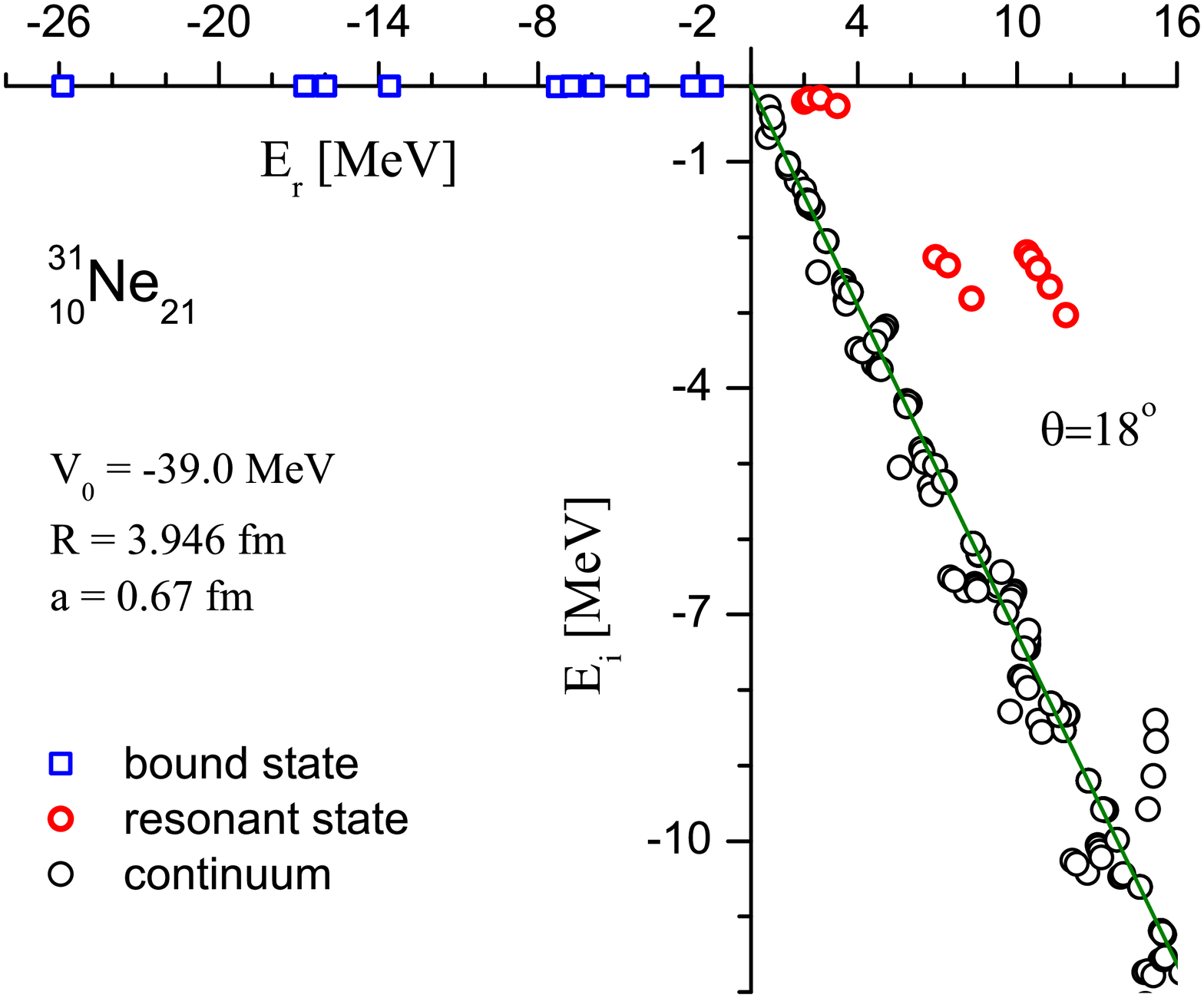}
\caption{ (Color online) The eigenvalues of the complex scaled Hamiltonian $H_{%
\protect\theta }$ for the states $\Omega^{\protect\pi}=\frac{1}{2}^{\pm},%
\frac{3}{2}^{\pm},\dots ,\frac{9}{2}^{\pm}$ in the present calculations with
the complex scaling parameter $\protect\theta =18^{\circ }$ and the
quadrupole deformation $\protect\beta_2=0.1$. The calculations are performed
by basis expansion with 60 HO shells. The bound states, the resonant states, and
the continuum are respectively labeled as open squares, red open circles,
and open circles. The solid line rotating with $2\protect\theta $ marks the
position of the continuous spectra.}
\end{figure}

\begin{figure}[tbp]
\includegraphics[width=8.6cm]{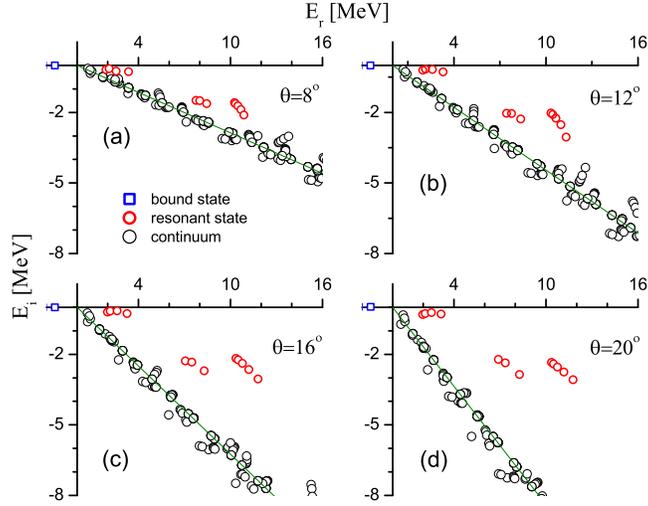}
\caption{ (Color online) The resonant and continuous spectra varying with
the complex rotation angle associated with the model. Except for the complex
rotation angle, the other parameters are the same as those in Fig.1.}
\end{figure}

\begin{figure}[tbp]
\includegraphics[width=8.6cm]{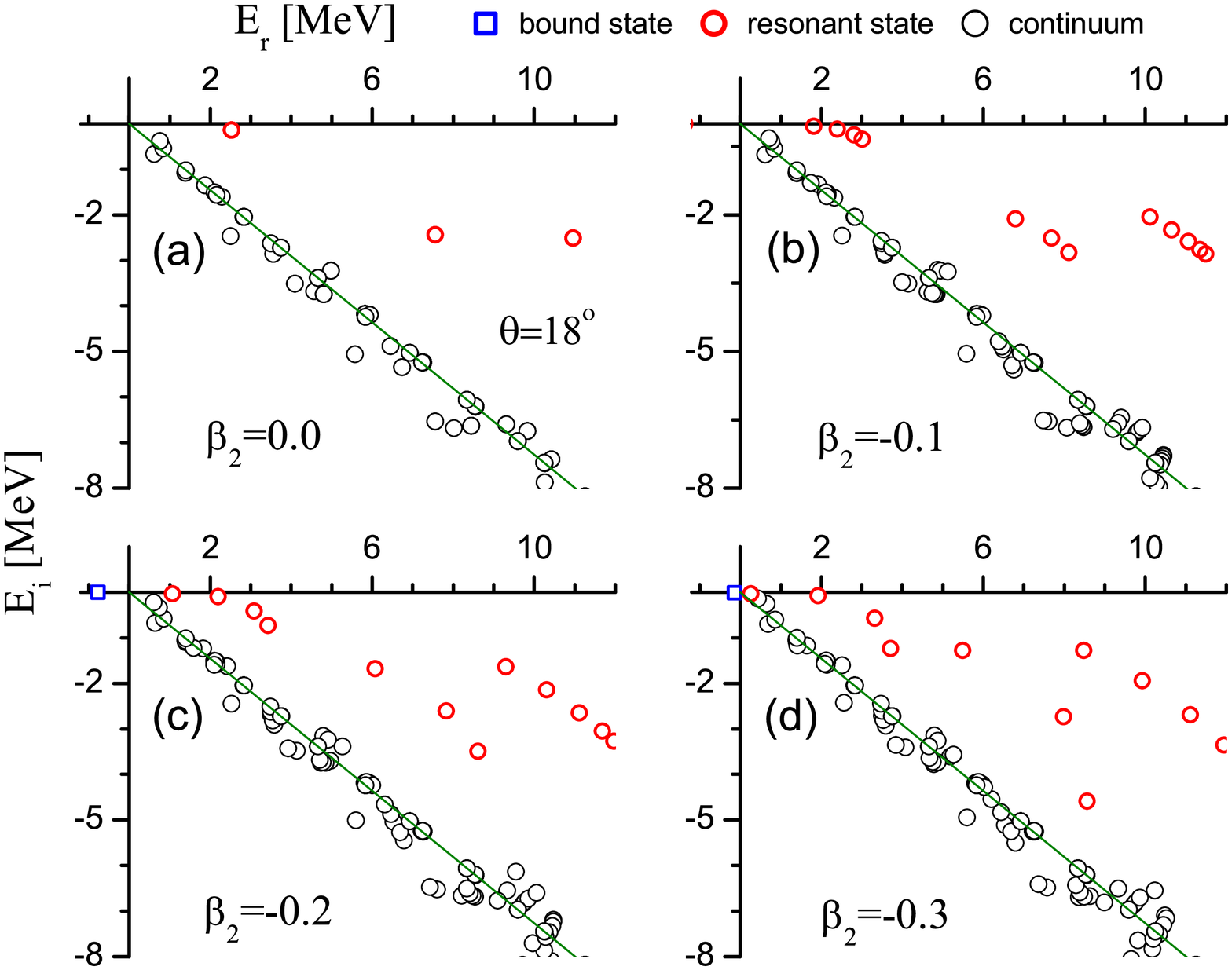}
\caption{ (Color online) The movement of the resonant states in the position
with deformation in the complex energy plane, where the development of
deformation is towards the direction of oblate shapes and the parameters
adopted are the same as those in Fig.1 except for a different deformation in
every configure.}
\end{figure}

\begin{figure}[tbp]
\includegraphics[width=8.cm]{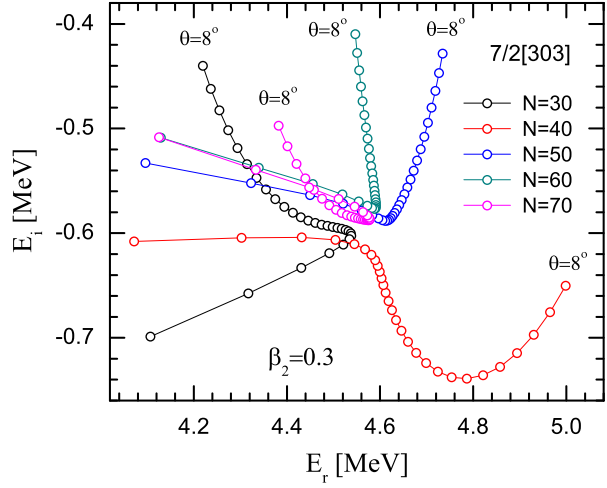}
\caption{ (Color online) The $\protect\theta$ trajectories corresponding to
the several different numbers of oscillator shells of the basis for the
resonant state 7/2[303] of the deformation point $\protect\beta_2=0.3$,
where $N$ is the quantum number of the main shell of the oscillator basis, and $%
\protect\theta $ varies from $8^{\circ }$ to $22^{\circ }$ by steps of $%
0.5^{\circ }$.}
\end{figure}

\begin{figure}[tbp]
\includegraphics[width=8.cm]{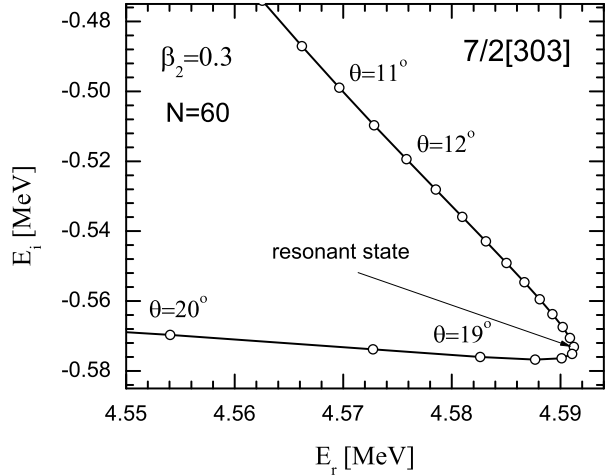}
\caption{ The same as Fig.4, but a close up of the $\protect\theta$
trajectory in Fig.4 for $N=60$, where the arrow marks the position of the
resonant state. }
\end{figure}
\begin{figure}[tbp]
\includegraphics[width=9cm]{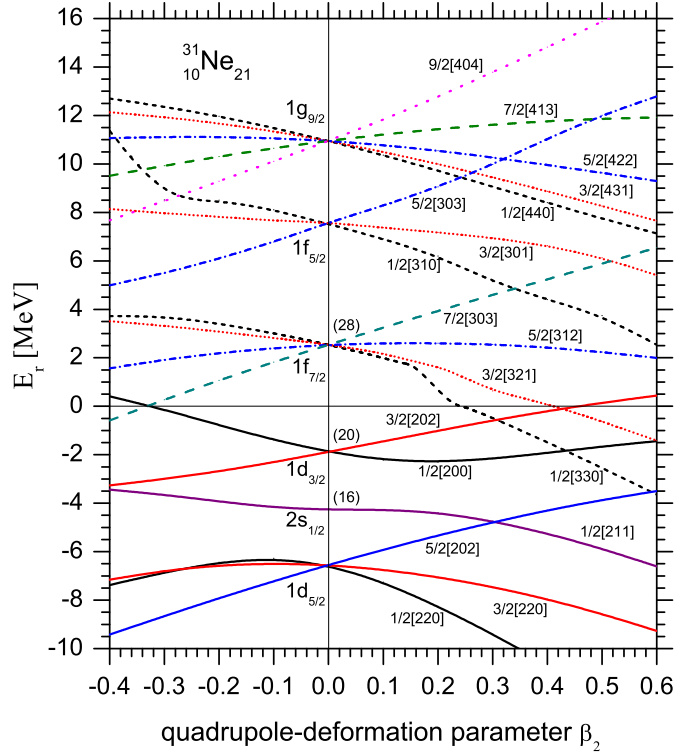}
\caption{ (Color online) Neutron single-particle levels in Woods-Saxon
potentials as a function of the quadrupole deformation parameter $\protect\beta%
_2 $. The depth, the diffuseness, and the radius of the potential are -39
MeV, 0.67 fm, and 3.946 fm, respectively. Every level is labeled with the
asymptotic quantum numbers $\Omega$[N$n_z\Lambda$].}
\end{figure}

\begin{figure}[tbp]
\includegraphics[width=8cm]{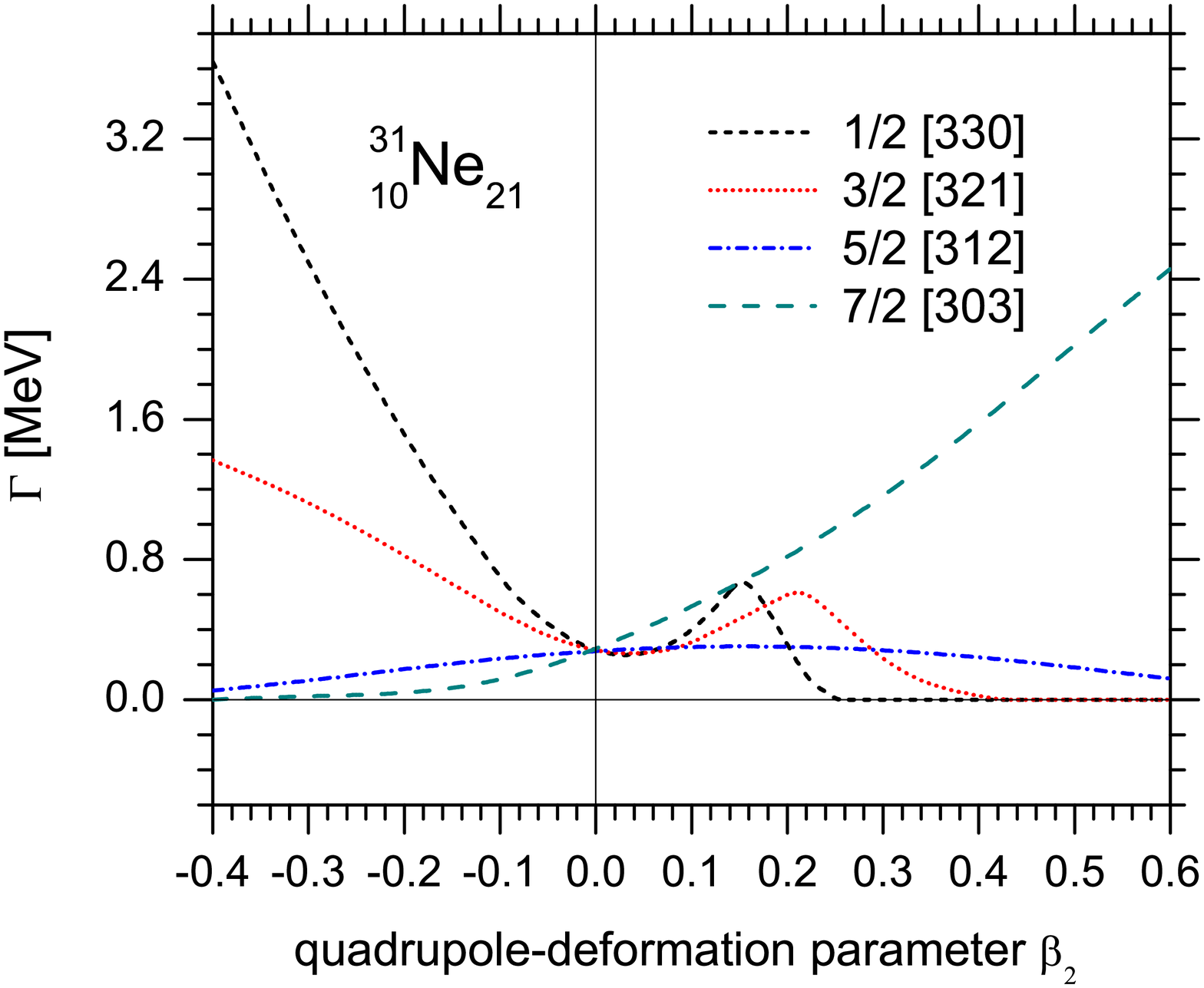}
\caption{ (Color online) The same as Fig.6, but for the width of the four
resonant states 1/2[330], 3/2[321], 5/2[312], and 7/2[303].}
\end{figure}


\begin{thebibliography}{99}
\bibitem{Dobaczewski96} J. Dobaczewski, W. Nazarewicz, T. R. Werner, J.-F.
Berger, C. R. Chinn, and J. Decharg\'e, Phys. Rev. C \textbf{53}, 2809
(1996).

\bibitem{Poeschl97} W. P\"oschl, D. Vretenar, G. A. Lalazissis, and P. Ring,
Phys. Rev. Lett. \textbf{79}, 3841 (1997).

\bibitem{Meng06} J. Meng, H. Toki, S.-G. Zhou, S. Q. Zhang, W. H. Long, and
L. S. Geng, Prog. Part. Nucl. Phys. \textbf{57}, 470 (2006).

\bibitem{Hamamoto10} I. Hamamoto Phys. Rev. C \textbf{81}, 021304(R) (2010).

\bibitem{Zhou10} S. G. Zhou, J. Meng, P. Ring, and E. G. Zhao, Phys. Rev. C
\textbf{82}, 011301(R) (2010).

\bibitem{Wigner47} E. Wigner and L. Eisenbud, Phys. Rev. \textbf{72}, 29
(1947).

\bibitem{Hale87} G. M. Hale, R. E. Brown, and N. Jarmie, Phys. Rev. Lett.
\textbf{59}, 763 (1987).

\bibitem{Humblet91} J. Humblet, B. W. Filippone, and S. E. Koonin, Phys.
Rev. C \textbf{44}, 2530 (1991).

\bibitem{Taylor72} J. R. Taylor, \textit{Scattering Theory: The Quantum
Theory on Nonrelativistic Collisions}, (John Wiley \& Sons, New York, 1972).

\bibitem{Hazi70} A. U. Hazi and H. S. Taylor, Phys. Rev. A \textbf{1}, 1109
(1970).

\bibitem{Ho83} Y. K. Ho, Phys. Rep. \textbf{99}, 1 (1983).

\bibitem{Kukulin89} V. I. Kukulin, V. M. Krasnopl'sky, and J. Hor\'{a}cek,
\textit{Theory of Resonances: Principles and Applications} (Kluwer Academic,
Dordrecht, 1989).

\bibitem{Taylor76} H. S. Taylor and A. U. Hazi, Phys. Rev. A \textbf{14},
2071 (1976).

\bibitem{Mandelshtam94} V. A. Mandelshtam, H. S. Taylor, V. Ryaboy, and N.
Moiseyev, Phys. Rev. A \textbf{50}, 2764 (1994).

\bibitem{Kruppa99} A. T. Kruppa and K. Arai, Phys. Rev. A \textbf{59}, 3556
(1999).

\bibitem{Zhang08} L. Zhang, S. G. Zhou, J. Meng, and E. G. Zhao, Phys. Rev.
C \textbf{77}, 014312 (2008).

\bibitem{Tanaka97} N. Tanaka, Y. Suzuki, and K. Varga, Phys. Rev. C \textbf{%
56}, 562 (1997).

\bibitem{Tanaka99} N. Tanaka, Y. Suzuki, K. Varga, and R. G. Lovas, Phys.
Rev. C \textbf{59}, 1391 (1999).

\bibitem{Yang01} S. C. Yang, J. Meng, and S. G. Zhou, Chin. Phys. Lett.
\textbf{18}, 196 (2001).

\bibitem{Zhang04} S. S. Zhang, J. Meng, S. G. Zhou, and G. C. Hillhouse,
Phys. Rev. C \textbf{70}, 034308 (2004).

\bibitem{Guo05} J.Y. Guo, R.D. Wang, and X.Z. Fang, Phys. Rev. C \textbf{72}%
, 054319 (2005).

\bibitem{Guo06} J.Y. Guo and X.Z. Fang, Phys. Rev. C \textbf{74}, 024320
(2006).

\bibitem{Moiseyev98} N. Moiseyev, Phys. Rep. \textbf{302}, 212 (1998).

\bibitem{Kruppa97} A. T. Kruppa, P. -H. Heenen, H. Flocard, and R. J.
Liotta, Phys. Rev. Lett. \textbf{79}, 2217 (1997).

\bibitem{Arai06} K. Arai, Phys. Rev. C \textbf{74}, 064311 (2006).

\bibitem{Aoyama06} S. Aoyama, T. Myo, K. Kat\={o}, and K. Ikeda, Prog.
Theor. Phys. \textbf{116}, 1 (2006).

\bibitem{Myo12} T. Myo, Y. Kikuchi, and K. Kato, Phys. Rev. C \textbf{85},
034338 (2012), and references therein.

\bibitem{Weder74} R. A. Weder, J. Math. Phys. \textbf{15}, 20 (1974).

\bibitem{Alhaidari07} A. D. Alhaidari, Phys. Rev. A \textbf{75}, 042707
(2007).

\bibitem{Bylicki08} M. Bylicki, G. Pestka, and J. Karwowski, Phys. Rev. A
\textbf{77}, 044501 (2008).

\bibitem{Guo10} J. Y. Guo, X. Z. Fang, P. Jiao, J. Wang, and B. M. Yao,
Phys. Rev. C \textbf{82}, 034318 (2010).

\bibitem{Nakamura09} T. Nakamura \textit{et al}., Phys. Rev. Lett. \textbf{%
103}, 262501 (2009).

\bibitem{Ferreira97} L. S. Ferreira, E. Maglione, and R. J. Liotta, Phys.
Rev. Lett. \textbf{78}, 1640 (1997).

\bibitem{Yoshida05} K. Yoshida and K. Hagino, Phys. Rev. C \textbf{72},
064311 (2005).

\bibitem{Cattapan00} G. Cattapan and E. Maglione, Phys. Rev. C \textbf{61},
067301(2000).

\bibitem{Hagino04} K. Hagino and N. Van Giai, Nucl. Phys. A \textbf{735}, 55
(2004).

\bibitem{Hagen06} G. Hagen and J. S. Vaagen, Phys. Rev. C \textbf{73},
034321 (2006).

\bibitem{Hamamoto05} I. Hamamoto, Phys. Rev. C \textbf{72}, 024301 (2005).

\bibitem{Li10} Z. P. Li, J. Meng, Y. Zhang, S. G. Zhou, and L. N. Savushkin,
Phys. Rev. C \textbf{81}, 034311 (2010).

\bibitem{Hamamoto12} I. Hamamoto, Phys. Rev. C \textbf{85}, 064329 (2012).

\bibitem{Aguilar71} J. Aguilar and J. M. Combes, Commun. Math. Phys. \textbf{%
22}, 269 (1971); E. Balslev and J. M. Combes, ibid. \textbf{22}, 280 (1971).
\end{thebibliography}
\end{document}